\documentclass[a4paper,11pt]{article}

\usepackage{jheppub} % for details on the use of the package, please
\usepackage[T1]{fontenc} % if needed

\usepackage{graphicx}
\usepackage{dcolumn}
\usepackage{bm}
\usepackage{amsmath}
\usepackage{amssymb}
\usepackage{makecell}
\usepackage{array}
\usepackage[utf8]{inputenc}

\newcommand{\beq}{\begin{equation}}
\newcommand{\eeq}{\end{equation}}
\newcommand{\bea}{\begin{eqnarray}}
\newcommand{\eea}{\end{eqnarray}}

\title{Flux tubes at finite temperature}

\author[a,b]{Paolo Cea}
\author[a]{Leonardo Cosmai}
\author[c]{Francesca Cuteri}
\author[c]{Alessandro Papa}

\affiliation[a]{INFN, Sezione di Bari,\\Via G. Amendola 173, I-70126 Bari, Italy}
\affiliation[b]{Dipartimento di Fisica, Universit\`a di Bari,\\Via G. Amendola 173, I-70126 Bari, Italy}
\affiliation[c]{Dipartimento di Fisica, Universit\`a della Calabria \& INFN-Cosenza,\\Ponte Bucci, cubo 31C, I-87036 Rende (Cosenza), Italy}

\emailAdd{paolo.cea@ba.infn.it}
\emailAdd{leonardo.cosmai@ba.infn.it}
\emailAdd{francesca.cuteri@cs.infn.it}
\emailAdd{papa@cs.infn.it}

\abstract{
The chromoelectric field generated by a static quark-antiquark pair, with
its peculiar tube-like shape, can be nicely described, at zero temperature, 
within the dual superconductor scenario for the QCD confining vacuum.
In this work we investigate, by lattice Monte Carlo simulations of the SU(3) 
pure gauge theory, the fate of chromoelectric flux tubes across the 
deconfinement transition. We find that, if the distance between the
static sources is kept fixed at about 0.76 fm $\simeq 1.6/\sqrt{\sigma}$ and
the temperature is increased towards and above the deconfinement temperature 
$T_c$, the amplitude of the field inside the flux tube gets smaller, while 
the shape of the flux tube does not vary appreciably across deconfinement.
This scenario with flux-tube ``evaporation'' above $T_c$ has no correspondence  
in ordinary (type-II) superconductivity, where instead the transition to the 
phase with normal conductivity is characterized by a divergent fattening of
flux tubes as the transition temperature is approached from below. 
We present also some evidence about the existence of flux-tube structures 
in the magnetic sector of the theory in the deconfined phase.
}

\begin{document}
\maketitle
\flushbottom

\section{Introduction}
\label{sect:intro}
Quarks and gluons, the elementary colored degrees of freedom of strong 
interactions, present some of the most interesting open issues within the 
Standard Model of particle physics.
In fact, strong interactions are described by Quantum ChromoDynamics (QCD), 
a local relativistic non-Abelian quantum field theory, which is not amenable 
to perturbation theory in the low-energy, large-distance regimes. 
However, many fundamental questions are linked to the large-scale
behavior of QCD. In particular, quarks and gluons appear to be confined in 
ordinary matter, due to the mechanism of color confinement which is not yet 
fully understood.
Reaching a detailed understanding of color confinement is one of the central 
goals of nonperturbative studies of QCD. \\
Lattice formulation of gauge theories allows us to investigate the color 
confinement phenomenon within a nonperturbative framework. Indeed, Monte Carlo 
simulations produce samples of vacuum configurations that, in principle, 
contain all the relevant information on the nonperturbative sector of QCD. \\
It is known since long that, in lattice numerical simulations,
tubelike structures emerge by analyzing the chromoelectric fields 
between static quarks~\cite{Fukugita:1983du,Kiskis:1984ru,Flower:1985gs,Wosiek:1987kx,DiGiacomo:1989yp,DiGiacomo:1990hc,Singh:1993jj,Cea:1992sd,Matsubara:1993nq,Cea:1992vx,Cea:1993pi,Cea:1994ed,Cea:1994aj,Cea:1995zt,Bali:1994de,Haymaker:2005py,D'Alessandro:2006ug,Cardaci:2010tb,Cea:2012qw,Cea:2013oba,Cea:2014uja,Cea:2014hma,Cardoso:2013lla,Caselle:2014eka}.
Such tubelike structures naturally lead to a linear potential between static 
color charges and, consequently, to a direct numerical evidence of color 
confinement~\cite{Bander:1980mu,Greensite:2003bk}. \\
Long time ago 't Hooft~\cite{'tHooft:1976ep} and 
Mandelstam~\cite{Mandelstam:1974pi} conjectured that the vacuum of QCD could be 
modeled as a coherent state of color magnetic monopoles, what is now known as 
a dual superconductor~\cite{Ripka:2003vv,Kondo:2014sta}. In the dual superconductor model of 
the QCD vacuum the condensation of color magnetic monopoles is analogous 
to the formation of Cooper pairs in the BCS theory of superconductivity. 
Remarkably, there are several numerical evidences~\cite{Shiba:1994db,Arasaki:1996sm,Cea:2000zr,Cea:2001an,DiGiacomo:1999fa,DiGiacomo:1999fb,Carmona:2001ja,Cea:2004ux,D'Alessandro:2010xg,Kato:2014nka} for the color magnetic condensation in 
QCD vacuum. However, it should be recognized~\cite{'tHooft:2004th} that the 
color magnetic monopole condensation in the confinement mode of QCD could be a 
consequence rather than the origin of the mechanism of color confinement,
that could actually arise from additional dynamical causes. 
Notwithstanding, the dual superconductivity picture of the QCD vacuum remains 
at least a very useful phenomenological frame to interpret the vacuum dynamics. 
\\
In previous studies~\cite{Cea:1992vx,Cea:1993pi,Cea:1994ed,Cea:1994aj,Cea:1995zt,Cardaci:2010tb,Cea:2012qw,Cea:2013oba,Cea:2014uja,Cea:2014hma} color flux tubes
made up of chromoelectric field directed along the line joining a static 
quark-antiquark pair have been investigated, in the cases of SU(2) and 
SU(3) pure gauge theories at zero temperature.
The aim of the present paper is to  extend the investigation of 
the structure of flux tubes  to  the case of the SU(3) pure gauge theory 
 {\em at finite temperatures}. 
In fact, on one hand the nonperturbative study in full QCD of the 
chromoelectric flux tubes generated by static color sources at finite 
temperature is directly relevant to clarify the formation of $c \bar{c}$ and 
$b \bar{b}$ bound states in heavy-ion collisions at high energies. On the other 
hand, the study of the behavior of the flux-tube parameters across the 
deconfining temperature in the SU(3) pure gauge theory allows us to check 
quantitatively the dual superconductor model of the QCD vacuum and to get 
hints about mechanisms possibly active also in full QCD. \\
The state of the art is the following. Differently from full QCD, which
exhibits a smooth crossover at about 170 MeV, the SU(3) pure gauge theory
undergoes a first order phase transition at $T_c \simeq 260$ MeV
(see Ref.~\cite{Boyd:1996bx} and references therein), separating a
low-temperature confined phase with a non-vanishing string tension from
the high-temperature deconfined phase with Debye-screened quark-antiquark 
potential and vanishing string tension. In the confined phase it has been
observed that, as the temperature approaches $T_c$, the string tension 
decreases, retaining however a non-zero values at 
$T_c$~\cite{Kaczmarek:1999mm,Cardoso:2011hh}. The interplay among the
string tension, which gives the energy per unit length in a (long enough) 
flux tube, the color fields, whose square contributes to the energy per 
unit volume, and the fields' spatial distribution, {\it i.e.} the shape of the 
flux tube, is not yet fully understood. There are, however, a few effective
descriptions, whose validity domain depends crucially on the distance
between the static sources, as nicely discussed in a recent 
paper~\cite{Baker:2015zlm}. For large enough distances (say, above 
$2/\sqrt{\sigma}$), the effective string theory 
approach~\cite{Luscher:1980ac,Luscher:1980fr,Luscher:1980iy} should hold, 
according to which the shape of the flux tube is determined by a fluctuating 
thin string connecting the sources. The implications of this approach on the
quark-antiquark potential and on the width of the flux tube have been studied 
numerically in SU($N$) gauge theories, both at $T=0$ and at $T<T_c$, in many 
papers~\cite{Caselle:2004er,Caselle:2005xy,Caselle:2006wr,Gliozzi:2010jh,Gliozzi:2010zv,Caselle:2011vk,Caselle:2012rp}. In several other
recent works~\cite{Cardoso:2013lla,Bakry:2011kga,Bakry:2014gea,Bakry:2012eq,Bakry:2015csa,Bakry:2011zz,Bakry:2010zt}, the detailed profile of the color field 
distribution near static sources has been studied, providing with information 
on the flux tube shape which goes well beyond the one encoded in its width.
The effective string theory approach is expected to fail at small distances and
close to $T_c$ even at large distances. At short distances between the sources,
the dual superconductivity picture should instead be valid, thus implying
that color fields between a quark-antiquark pair can be described in the same fashion as 
isolated vortex solutions in ordinary superconductors. It would be
extremely interesting to study by numerical Monte Carlo simulations the shape 
of flux tubes in a wide enough range of distances between the static sources 
and for various temperatures around $T_c$ to cover both domains where the 
dual superconductivity picture and the effective string theory approach are 
expected to hold. 

As a first step in this direction, in this paper we study the
profile of flux tubes at a fixed distance, about 0.76 fm, corresponding
to about $1.6/\sqrt{\sigma}$, scanning the temperature in the range 
$0.8\,T_c \div T_c$, with the aim of understanding
the mechanism underlying the lowering of the string tension, {\it i.e.} if
it is dominated by the weakening of the color fields in the flux tube or
by the broadening of the flux tube itself. Moreover, we extend our analysis 
also to temperatures in the range $T_c \div 1.2\, T_c$, {\it i.e.} in the domain
of color charge screening, to see how the expected vanishing of the string 
tension in this region reflects in the shape of flux tubes. The part
of investigation in the latter temperature domain shoud be understood
just as ``numerical experiment'', without any prejudice about possible
results and their explanation.

To implement this program, however, we need to perform numerical simulations
on lattices with very large volumes. To this end, we have made use of the 
publicly available MILC code~\cite{MILC}, which has been suitably modified by 
us in order to introduce the relevant observables.  
Indeed, the use of the MILC code will permit to do simulations for the 
physically relevant case of full QCD with dynamical quarks. \\
The plan of the paper is as follows. In Section~\ref{setup} we discuss
the observables needed to extract the field strength tensor of the
static quark-antiquark sources and  present some consistency checks
of our code. Section~\ref{results} is devoted to the discussion of 
finite-temperature results. In particular we critically analyze the behavior  
of the coherence and penetration lengths across the deconfining transition.
In Section~\ref{magnetic} we discuss the structure of the flux tubes in the 
magnetic sector at finite temperature, also  in the deconfined phase. 
Finally, in Section~\ref{conclusions},
we summarize our results and present our conclusions. 
\section{Lattice setup}
\label{setup}
\begin{figure}[htb] 
\centering
\includegraphics[scale=0.7,clip]{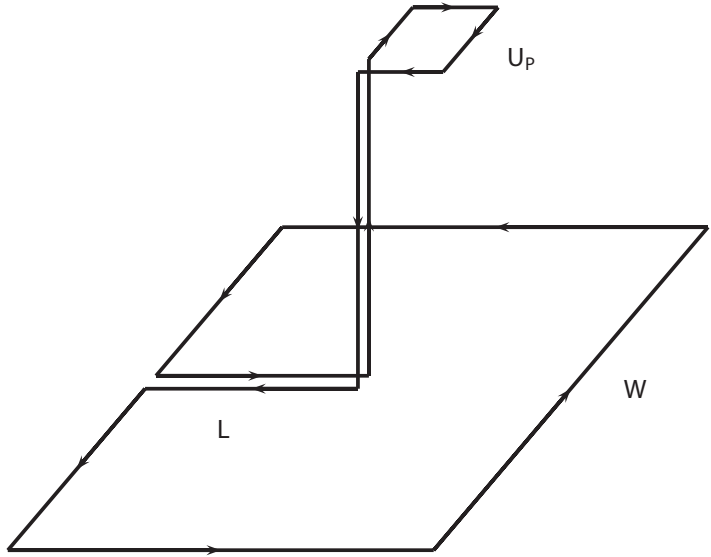} 
\caption{The connected correlator given in Eq.~(\protect\ref{rhoW})
between the plaquette $U_{P}$ and the Wilson loop
(subtraction in $\rho_{W}^{\rm conn}$ not explicitly drawn).}
\label{fig:op_W}
\end{figure}

To explore on the lattice the field configurations produced by 
a static quark-antiquark pair, the following connected correlation 
function~\cite{DiGiacomo:1989yp,DiGiacomo:1990hc,Kuzmenko:2000bq,
DiGiacomo:2000va} was used:
\begin{equation}
\label{rhoW}
\rho_W^{\rm conn} = \frac{\left\langle {\rm tr}
\left( W L U_P L^{\dagger} \right)  \right\rangle}
              { \left\langle {\rm tr} (W) \right\rangle }
 - \frac{1}{N} \,
\frac{\left\langle {\rm tr} (U_P) {\rm tr} (W)  \right\rangle}
              { \left\langle {\rm tr} (W) \right\rangle } \; ,
\end{equation}
where $U_P=U_{\mu\nu}(x)$ is the plaquette in the $(\mu,\nu)$ plane, connected
to the Wilson loop $W$ by a Schwinger line $L$, and $N$ is the number of colors
(see Fig.~\ref{fig:op_W}).
The correlation function defined in Eq.~(\ref{rhoW}) measures the field 
strength, since in the naive continuum limit~\cite{DiGiacomo:1990hc,DelDebbio:1994zn}
\begin{equation}
\label{rhoWlimcont}
\rho_W^{\rm conn}\stackrel{a \rightarrow 0}{\longrightarrow} a^2 g 
\left[ \left\langle
F_{\mu\nu}\right\rangle_{q\bar{q}} - \left\langle F_{\mu\nu}
\right\rangle_0 \right]  \;,
\end{equation}
where $\langle\quad\rangle_{q \bar q}$ denotes the average in the presence of 
a static $q \bar q$ pair and $\langle\quad\rangle_0$ is the vacuum average,
which is expected to vanish. Accordingly, we are led to define the 
quark-antiquark field strength tensor as:
\begin{equation}
\label{fieldstrengthW}
F_{\mu\nu}(x) = \sqrt\frac{\beta}{2 N} \, \rho_W^{\rm conn}(x)   \; .
\end{equation}
To specify better the color structure of the field $F_{\mu\nu}$, we note that
the Wilson loop connected to the plaquette is the source of a color field
which points, in average, onto an unknown direction $n^a$ in color space,
given by the loop itself (there is no preferred direction). What we measure is 
the average projection of the color field onto that direction. The color 
indices of the Schwinger lines are contracted with the loop, which is the 
source of the field, and realize the color parallel transport between the 
source loop and the plaquette position. For this reason, the $F_{\mu\nu}$
appearing in Eq.~(\ref{fieldstrengthW}), should be understood as 
$n^a F_{\mu\nu}^a$,
\begin{equation}
\rho_W^{\rm conn}\stackrel{a \rightarrow 0}{\longrightarrow} a^2 g
\left[ \left\langle
n^aF^a_{\mu\nu}\right\rangle_{q\bar{q}} \right]\;.
\end{equation}
That this relation must hold and that the vector in color space 
$n^a$ must be introduced follows from the linearity in the color field
of the operator in~(\ref{rhoWlimcont}) and from its gauge invariance.
Similar considerations apply to the operator~(\ref{eq:rhopconn}) defined 
below (see also Ref.~\cite{Cea:2015mqi}).
In Eq.~(\ref{rhoW}) the Schwinger line $L$ is taken in such a way that it leaves
the plane of the Wilson loop just at the center of the latter, which means
that the field is measured on the plane cutting orthogonally the line between 
the static sources in its mid-point. We have not considered here other
possibilities, thus implying that the tubular shape of the flux profile is only
assumed.

In the dual superconductor model of the QCD vacuum, the formation of the 
chromoelectric flux tube can be interpreted as the dual Meissner effect. 
In this context the transverse shape of the longitudinal chromoelectric field 
$E_l$ should resemble the dual version of the Abrikosov vortex field 
distribution. Therefore, the proposal was 
advanced~\cite{Cea:1992sd,Cea:1992vx,Cea:1993pi,Cea:1994ed,Cea:1994aj,
Cea:1995zt} to fit the transverse shape of the longitudinal chromoelectric 
field according to
\begin{equation}
\label{London}
E_l(x_t) = \frac{\phi}{2 \pi} \mu^2 K_0(\mu x_t) \;,\;\;\;\;\; x_t > 0 \; ,
\end{equation}
where $K_n$ is the modified Bessel function of order $n$, $\phi$ is
the external flux, and $\lambda=1/\mu$ is the London penetration length. 
Note that Eq.~(\ref{London}) is valid as long as $\lambda \gg \xi$, 
$\xi$ being the coherence length (type-II superconductor), which measures the 
coherence of the magnetic monopole condensate (the dual version of the Cooper 
condensate). 

However, several numerical studies~\cite{Suzuki:1988yq,Maedan:1989ju,
Singh:1993ma,Singh:1993jj,Matsubara:1994nq,Schlichter:1997hw,Bali:1997cp,
Schilling:1998gz,Gubarev:1999yp,Koma:2001ut,Koma:2003hv}, in both SU(2) and 
SU(3) lattice gauge theories, have shown that the confining vacuum seems to 
behave like an effective dual superconductor which lies on the borderline 
between a type-I and a type-II superconductor. If this is the case, 
Eq.~(\ref{London}) is no longer adequate to account for the transverse 
structure of the longitudinal chromoelectric field. In fact, 
in Refs.~\cite{Cea:2012qw,Cea:2013oba,Cea:2014uja,Cea:2014hma} it has been 
suggested that lattice data for chromoelectric flux tubes can be analyzed by 
exploiting the results presented in Ref.~\cite{Clem:1975aa}, where, from the 
assumption of a simple variational model for the magnitude of the normalized 
order parameter of an isolated vortex, an analytic expression is derived for 
magnetic field and supercurrent density, that solves the Ampere's law and the 
Ginzburg-Landau equations. As a consequence, the transverse distribution of  
the chromoelectric flux tube can be  described, according 
to~\cite{Cea:2012qw,Cea:2013oba,Cea:2014uja,Cea:2014hma}, by
\begin{equation}
\label{clem1}
E_l(x_t) = \frac{\phi}{2 \pi} \frac{1}{\lambda \xi_v} \frac{K_0(R/\lambda)}
{K_1(\xi_v/\lambda)} \; ,
\end{equation}
where
\begin{equation}
\label{rrr}
 R=\sqrt{x_t^2+\xi_v^2}
\end{equation}
and $\xi_v$ is a variational core-radius parameter.
Equation~(\ref{clem1}) can be rewritten as
\begin{equation}
\label{clem2}
E_l(x_t) =  \frac{\phi}{2 \pi} \frac{\mu^2}{\alpha} \frac{K_0[(\mu^2 x_t^2 
+ \alpha^2)^{1/2}]}{K_1[\alpha]} \; ,
\end{equation}
with
\begin{equation}
\label{alpha}
\mu= \frac{1}{\lambda} \,, \quad \frac{1}{\alpha} =  \frac{\lambda}{\xi_v} \,.
\end{equation}
By fitting Eq.~(\ref{clem2}) to flux-tube data, one can get 
both the penetration length $\lambda$ and the ratio of the penetration length 
to the variational core-radius parameter, $\lambda/\xi_v$. Moreover,   
the Ginzburg-Landau $\kappa$ parameter can be obtained by
\begin{equation}
\label{landaukappa}
\kappa = \frac{\lambda}{\xi} =  \frac{\sqrt{2}}{\alpha} 
\left[ 1 - K_0^2(\alpha) / K_1^2(\alpha) \right]^{1/2} \,.
\end{equation}
Finally, the coherence length $\xi$ is determined by combining  
Eqs.~(\ref{alpha}) and~(\ref{landaukappa}).

We have already said that our aim is to extend previous studies of the 
structure of flux tubes performed at zero temperature to the case of SU(3) 
pure gauge theory at finite temperatures. From the phenomenological point of 
view, the nonperturbative study of the chromoelectric flux tubes generated by 
static color sources at finite temperature is directly relevant to clarify the 
formation of $c \bar{c}$ and $b \bar{b}$ bound states in heavy-ion  
collisions at high energies.  It should be evident, however, that to implement 
this program we cannot employ the Wilson loop operator in the connected 
correlation in Eq.~(\ref{rhoW}). This problem can be easily overcome if we 
replace in Eq.~(\ref{rhoW}) the Wilson loop with two Polyakov lines.
In addition, we also need to surrogate the cooling mechanism previously 
used to enhance the signal-to-noise ratio.
Indeed, cooling is a well established method for locally 
suppressing quantum fluctuations in gauge field configurations. 
However, at finite temperatures the cooling procedure tends to 
suppress also thermal fluctuations.
Fortunately, there is an alternative, yet somewhat related, approach that
is the application of APE smearing~\cite{Falcioni1985624,Albanese1987163}
to the gauge field configurations. This approach also leads to the desirable 
effect of suppressing lattice artifacts at the scale of the cutoff,
without affecting the thermal fluctuations. Moreover, this procedure can 
be iterated many times to obtain smoother and smoother gauge field 
configurations and allows the anisotropic treatment of spatial and
temporal links.\\
\begin{figure}[tb] 
\centering
\includegraphics[scale=0.3,clip]{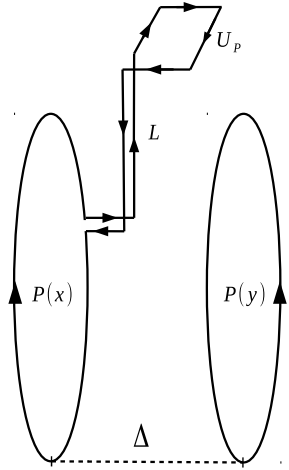} 
\caption{The connected correlator given in Eq.~(\protect\ref{eq:rhopconn})
between the plaquette $U_{P}$ and the Polyakov loops 
(subtraction in $\rho_{P}^{\rm conn}$ not explicitly drawn).}
\label{fig:op_P}
\end{figure}
In fact, in Ref.~\cite{Cea:2013oba,Cea:2014uja,Cea:2014hma} we suggested that  
the following connected correlations (depicted in Fig.~\ref{fig:op_P}):  
\bea
\label{eq:rhopconn}
\rho_{P}^{\rm conn}&=&\frac{\left\langle \mathrm{tr}\left(P\left(x\right)LU_{P}
L^{\dagger}\right)\mathrm{tr}P^\dagger\left(y\right)\right\rangle }{\left\langle 
\mathrm{tr}\left(P\left(x\right)\right)\mathrm{tr}\left(P^\dagger\left(y\right)\right)
\right\rangle } \\
&-&\frac{1}{3}\frac{\left\langle \mathrm{tr}\left(P\left(x\right)\right)
\mathrm{tr}\left(P^\dagger\left(y\right)\right)\mathrm{tr}\left(U_{P}\right)\right
\rangle }{\left\langle \mathrm{tr}\left(P\left(x\right)\right)\mathrm{tr}
\left(P^\dagger\left(y\right)\right)\right\rangle}\; , \nonumber
\eea
where the two Polyakov lines are separated by a distance $\Delta$, could 
replace the correlator with the Wilson loop defined in Eq.~(\ref{rhoW}).  
Even in this case, after taking into account Eqs.~(\ref{rhoWlimcont}) 
and~(\ref{fieldstrengthW}), we may define the field strength tensor as:
\begin{equation}
\label{fieldstrengthP}
 F_{\mu\nu}\left(x\right)=\sqrt{\frac{\beta}{6}}\rho_{P}^{\rm conn}
\left(x\right).
\end{equation}
A detailed derivation of Eq.~(\ref{fieldstrengthP}), together with the
discussion of its physical interpretation, can be found in 
Ref.~\cite{Skala:1996ar}. \\
Obviously, one must preliminarily check that this method gives results
which are consistent with previous studies obtained with Wilson loops
and cooling. In fact, in Refs.~\cite{Cea:2013oba,Cea:2014uja,Cea:2014hma}
we showed that results obtained with the operator Eq.~(\ref{eq:rhopconn})
are consistent within statistical uncertainties with the results obtained by 
employing Wilson loops and the cooling procedure. In~\cite{Cea:2014uja}
we discussed also the comparison with the approach of 
Ref.~\cite{Bicudo:2014wka}, where a disconnected correlator of plaquette and 
Wilson loop was adopted, and  with  that of Ref.~\cite{Caselle:2012rp},
where the adopted probe observable was the disconnected correlator of two
Polyakov lines and a plaquette. \\
\begin{figure}[tb] 
\centering
\includegraphics[scale=0.6,clip]{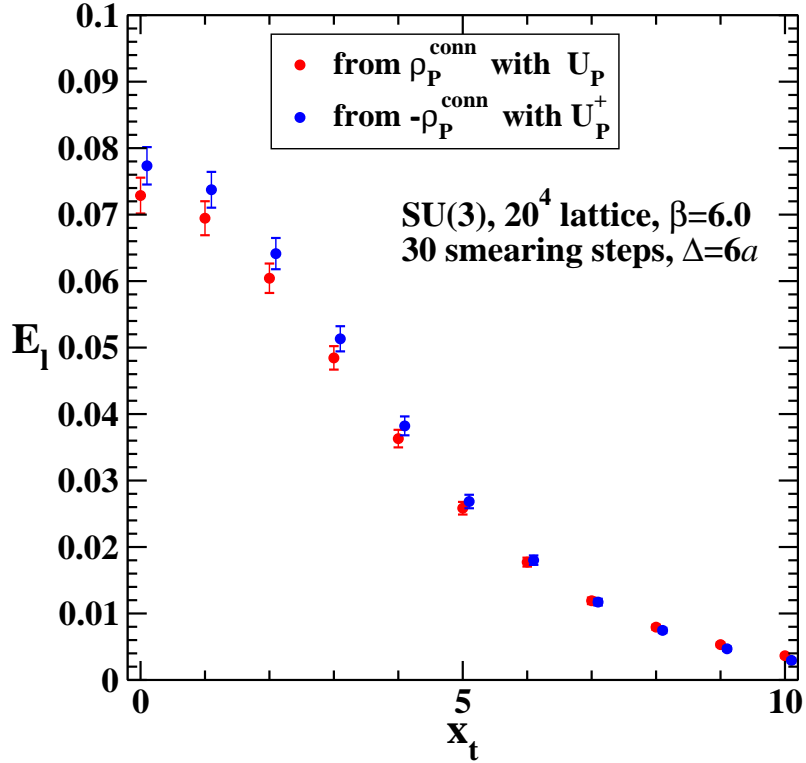}
\caption{(color online). Longitudinal chromoelectric field for the two different
orientations of the plaquette entering the definition of the connected 
correlator given in Eq.~(\protect\ref{eq:rhopconn}).}
\label{fig:lin}
\end{figure}
Our lattice setup is as follows.
In order to reduce the ultraviolet noise, we applied to the operator in 
Eq.~(\ref{eq:rhopconn}) one step of HYP smearing~\cite{Hasenfratz:2001hp} 
to temporal links, with smearing parameters $(\alpha_1,\alpha_2,\alpha_3) 
= (1.0, 0.5, 0.5)$, and $N_{\rm APE}$ steps of APE 
smearing~\cite{Falcioni1985624,Albanese1987163} to spatial links, with 
smearing parameter $\alpha_{\rm APE} = 0.50$. Here $\alpha_{\rm APE}$ is the ratio 
between the weight of one staple  and  the weight of the original link.
We expect that the smearing procedure should improve the approach to the
continuum and, therefore, enforce the validity of the continuum limit
given in Eq.~(\ref{rhoWlimcont}).
We performed numerical simulations using the Wilson action on lattices
with periodic boundary conditions and the heat-bath algorithm combined with 
overrelaxation. For each value of the gauge coupling we collected
4000 - 5000 sweeps; one sweep corresponds to four overrelaxation steps 
followed by one heat-bath step.
To allow thermalization we typically discarded  a few thousand  sweeps and, 
in order to reduce the autocorrelation time, measurements were taken after 10
updatings. The error analysis was performed by the jackknife method over 
bins at different blocking levels.   \\
Since we are adopting a new numerical code build from the MILC code, we have, 
preliminarily, performed some consistency checks. First, we simulated the 
SU(3) pure gauge theory a zero temperature on larger lattices.
We performed numerical simulations on $32^4$ lattices and measured the operator 
given in Eq.~(\ref{rhoW}).
In fact, we obtained results for the field strength tensor which, within the statistical uncertainties, were compatible with the ones obtained
in Refs.~\cite{Cea:2013oba,Cea:2014uja,Cea:2014hma} on $20^4$ lattices.  
After that, we checked that our operator is sensitive 
to the field strength tensor and not to its square. To this end, it is 
enough to check that:
\begin{equation}
\label{strength}
F_{\mu\nu}(x) \; = \; - \;  F_{\nu\mu}(x)    \; ,
\end{equation}
where $F_{\mu\nu}(x)$ is defined by Eq.~(\ref{fieldstrengthP}).  This amounts
to change $U_P$ into $U_P^\dagger$  in Eq.~(\ref{eq:rhopconn}), for a fixed 
choice of the $\mu\nu$-plane where the plaquette lies.  
In fact, in Fig.~3 it is shown that, under this transformation, two independent simulations, differing
in the choice of the orientation of the plaquette, give two field strengths 
with opposite sign, within 1$\sigma$ accuracy.
The small deviations seen in Fig.~\ref{fig:lin} from the exact linearity
can be attributed to lattice artifacts and are expected to vanish in the
$a\to 0$ limit.
Finally, we have checked that also at finite temperatures only the longitudinal 
chromoelectric field gave a statistically sizable signal. Therefore, in 
the following, we will focus only on the numerical results regarding the 
longitudinal chromoelectric fields.
\section{Numerical results}
\label{results}
We performed numerical simulations at finite temperatures on lattices with
temporal extension ranging from $L_t=10$  up to $L_t=16$ and spatial
size $L_s$ fixed as to have aspect ratio $L_s/L_t \ge 4$. The temperature 
of the gauge system is varied according to
\begin{equation}
\label{temperature}
T \; = \; \frac{1}{a(\beta) \, L_t} \; ,
\end{equation}
where the scale is fixed using the parameterization~\cite{Edwards:1998xf}:
\bea
\label{sqrt-sigma-SU3}
\left ( a \, \sqrt{\sigma} \right )(g) &=& f_{{\rm{SU(3)}}}(g^2) 
\left \{ 1+0.2731\,\hat{a}^2(g)  \right .\\
&-&0.01545\,\hat{a}^4(g) +0.01975\,\hat{a}^6(g) \left . \right \}/ 0.01364 \;  , \nonumber
\eea
\[
\hat{a}(g) = \frac{f_{{\rm{SU(3)}}}(g^2)}{f_{{\rm{SU(3)}}}(g^2(\beta=6))} 
\;, \;
\beta=\frac{6}{g^2} \,, \;\;\; 5.6 \leq \beta \leq 6.5\;,
\]
with
\beq
\label{fsun}
f_{{\rm{SU(3)}}}(g^2) = \left( {b_0 g^2}\right)^{- b_1/2b_0^2} 
\, \exp \left( - \frac{1}{2 b_0 g^2} \right) \,,
\eeq
\[
b_0 \, = \, \frac{11}{(4\pi)^2} \; \; , \; \; b_1 \, = \, \frac{102}{(4\pi)^4} 
\; .
\] \; 
In the following, we assumed for the string tension the standard value 
of $\sqrt{\sigma} = 420$ MeV. \\
\begin{figure}[tb] 
\centering
\includegraphics[scale=0.6,clip]{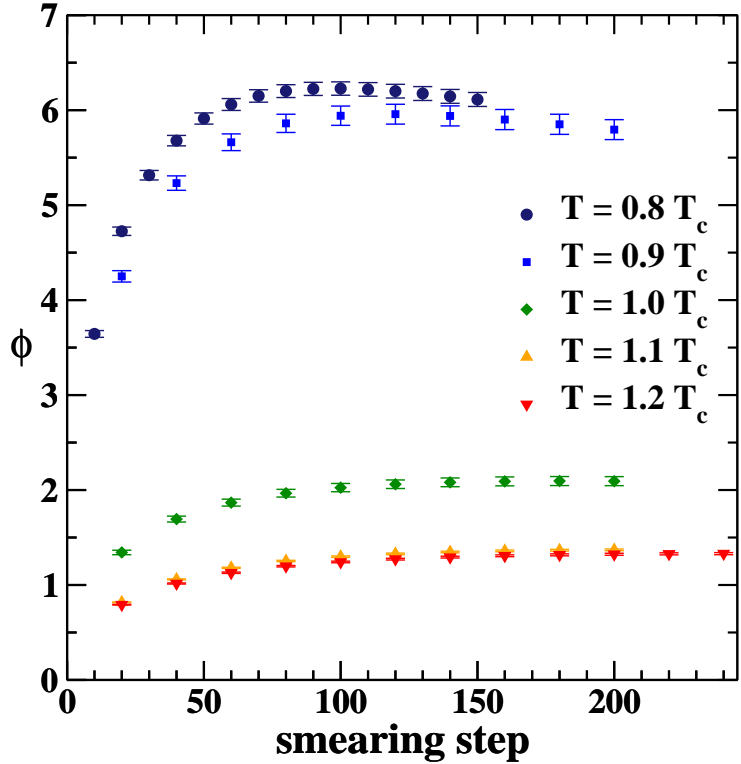}
\caption{(color online). Behavior of the parameter $\phi$ {\it vs} smearing on 
a $40^3\times 10$ lattice from measurements of the Polyakov connected correlator at 
different nonzero temperatures.}
\label{fig:phi_vs_smear}
\end{figure}
We measured the connected correlator given in Eq.~(\ref{eq:rhopconn}) at 
the middle of the line connecting the static color sources, for various 
values of the distance between the sources and for integer transverse 
distances. As already discussed, to reduce statistical fluctuations in gauge 
field configurations, we performed measurements after several APE smearing 
steps. At each smearing step, we fitted our data for the transverse 
shape of the longitudinal chromoelectric field to Eq.~(\ref{clem2}).  
Remarkably, we found that Eq.~(\ref{clem2}) is able to reproduce the transverse 
profile of the longitudinal chromoelectric field even at finite temperatures.
As a result, we obtained the fit parameters for different smearing steps. This 
allowed us to check the dependence of these parameters on the number of 
smearing steps. To fix the optimal value of the smearing step, we looked at
well defined plateaux in the values of the fit parameters {\it versus} 
the smearing step. We found that the most reasonable choice was to 
look at plateaux for the parameter $\phi$, which is related to the flux of the 
chromoelectric field: since this observable encodes both the amplitude of 
the chromoelectric field and the transverse size of the flux tube, it is 
the best candidate to indicate the disentanglement of the signal from 
the background noise.
In Fig.~\ref{fig:phi_vs_smear} we display the fitted parameter $\phi$ 
{\it versus} the smearing steps for the different temperatures considered in 
this paper.
We see that, indeed, $\phi$ displays rather shallow plateaux at 
$N_{\rm APE} \sim 100$ for all the adopted temperatures.
%%%%%%%%%%%%%%%%%%%%%%%%%%%%%
\begin{table}[thb]
\setlength{\tabcolsep}{0.2cm}
\begin{center} 
\caption{Summary of simulation parameters used to check the scaling of the 
longitudinal chromoelectric field.}
\label{tab:run_summary}
\begin{tabular}{|c|c|c|c|c|>{\centering\arraybackslash}m{1.5cm}|}
\hline\hline
$\beta$ & $L_s\times L_t$ & $\Delta$ [fm] & $T/T_c$ & statistics & optimal 
smearing step
 \\ \hline
  6.05  & $40^3\times 10$ & 0.714 & 0.8 & 3000 & 100\\
%         & $32^3\times 8$  &      &             &                \\
%   6.17  & $40^3\times 10$ &      &             &                \\
%         & $48^3\times 12$ &      &             &                \\ \hline
  6.37  & $64^3\times 16$ & 0.761 & 0.8 & 4000 & 180\\
\hline \hline 
\end{tabular}
\end{center}
\end{table}
\begin{figure}[tb] 
\centering
\includegraphics[scale=0.6,clip]{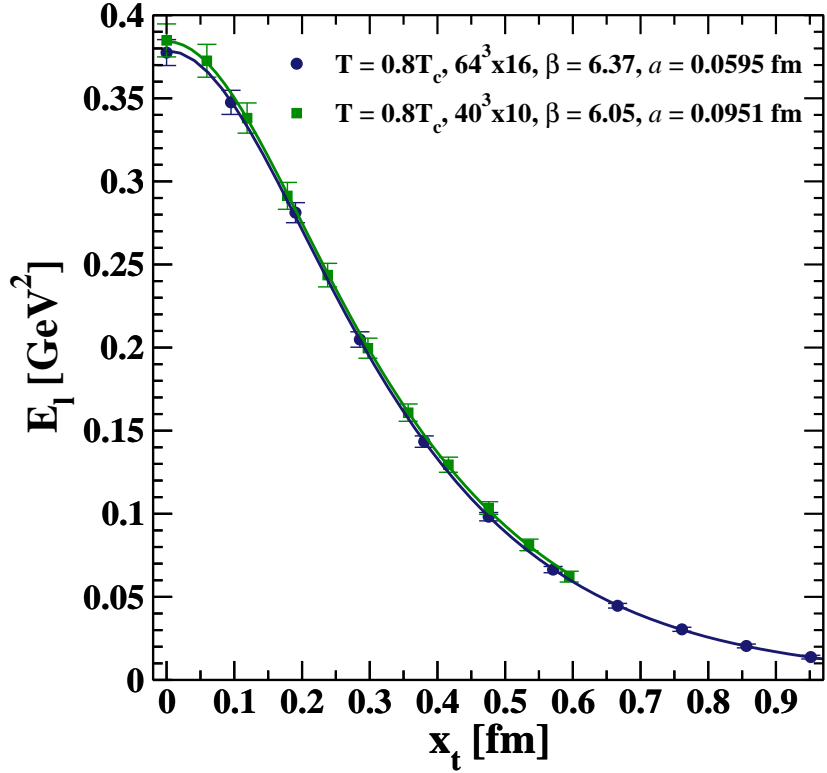} 
\caption{(color online). Comparison of the behavior of the longitudinal 
chromoelectric field at a fixed temperature, $T\simeq 0.8\, T_c$, obtained by 
different combinations of $\beta$ and $L_t$, as showed in 
Table~\ref{tab:run_summary}. The solid lines are the fit of our data to 
Eq.~(\ref{clem2}).}
\label{fig:E_vs_beta}
\end{figure}
We looked also for contamination effects on the longitudinal chromoelectric 
field due to the presence of the static color sources. 
To this aim, we varied the distance $\Delta$ between the Polyakov lines
keeping the temperature fixed at $T/T_c = 0.8$ and measured the longitudinal 
chromoelectric field on lattices with different values of $L_t$ and of the
gauge coupling $\beta$. This allowed us to size the cut-off effects and to
single out the scaling region in $\beta$.
The results of our study showed that fixing the distance between the 
static color sources such that $\Delta \gtrsim 0.7$ fm was a good compromise 
between the absence of spurious contamination effects due to the static
color sources and a reasonable signal-to-noise ratio. In addition, we found 
that the longitudinal chromoelectric field displays a nice scaling behavior if 
one adopts lattices with $L_t \ge 10$. In fact, in Fig.~\ref{fig:E_vs_beta} we 
compare the transverse profile of the longitudinal chromoelectric field for two 
different lattice setups, as summarized in Table~\ref{tab:run_summary}.
From Fig.~\ref{fig:E_vs_beta}, we see that the chromoelectric field seems to 
display an almost perfect scaling. \\
Having selected the gauge coupling region where continuum scaling holds, 
we focused on the temperature dependence of the longitudinal 
chromoelectric field. We measured the connected correlator given in
Eq.~(\ref{eq:rhopconn}) on $40^3\times 10$ lattices for physical temperatures 
ranging from $0.8 \, T_c$ up to $1.2 \, T_c$.
We chose the distance $\Delta$ between the two Polyakov lines around
0.76 fm. In Table~\ref{tab:fit_summary} we summarize the simulation 
setup and the corresponding best-fit values of the parameters.
\begin{table}[thb]
\begin{center} 
\caption{Simulation parameters for the lattice 
$L_s\times L_t = 40^3\times 10$, fitted values of the parameters, and reduced 
chi-square (chromoelectric sector).}
\label{tab:fit_summary}
\begin{tabular}{|c|c|c|c|c|c|c|c|}
\hline\hline
$\beta$ & $\Delta$ [fm] & $T/T_c$ & $\phi$ & $\mu$ & $\xi_v$ & $\chi_r^2$ \\ \hline
6.050	&0.761	&0.8	&6.201(68)	&0.382(13)	&3.117(191)	&0.02\\
6.125	&0.761	&0.9	&5.941(101)	&0.337(20)	&3.652(360)	&0.01\\
6.200	&0.756	&1.0	&2.061(45)	&0.328(22)	&3.312(389)	&0.01\\
6.265	&0.757	&1.1	&1.359(9)	&0.344(7)	&4.286(131)	&0.06\\
6.325	&0.760	&1.2	&1.324(11)	&0.332(8)	&4.248(142)	&0.06\\
\hline\hline 
\end{tabular} 
\end{center}
\end{table}
\begin{figure}[tb] 
\centering
\includegraphics[scale=0.6,clip]{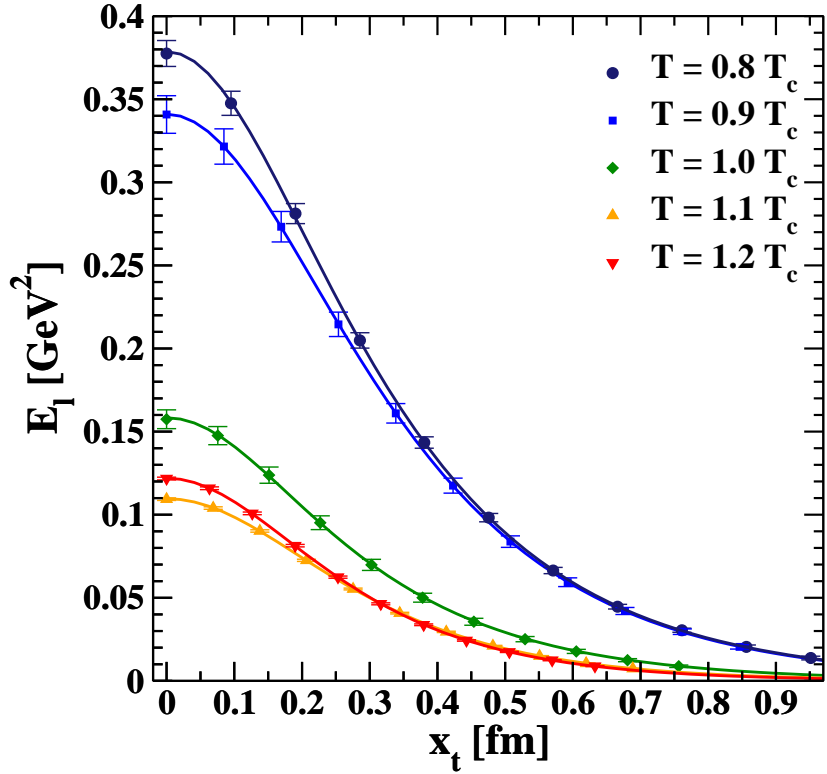} 
\caption{(color online). Behavior of the longitudinal chromoelectric field at 
a fixed lattice size $40^3\times 10$ and various gauge couplings in the 
scaling region {\it vs} the transverse distance. 
The solid lines are the fit of our data to Eq.~(\ref{clem2}).}
\label{fig:E_vs_T}
\end{figure}
In Fig.~\ref{fig:E_vs_T} we display the transverse distribution of the 
longitudinal chromoelectric field for the different temperatures
used in the present study. From Fig.~\ref{fig:E_vs_T} we infer that, as the 
temperature is increased towards and above the deconfinement temperature $T_c$, 
the strength of the flux-tube chromoelectric field decreases very quickly,  
while the size of the flux tube does not seem to vary appreciably. 
This behavior suggests that, by increasing the temperature above the 
critical one, the flux tube is evaporating while almost preserving his shape.
This scenario with flux-tube evaporation above $T_c$ has no correspondence 
in ordinary type-II superconductivity, where instead the transition to the 
phase with normal conductivity is characterized by a divergent fattening of
flux tubes as the transition temperature is approached from below. 
A difference in the behavior is to be expected, given that the transition in 
SU(3) is first order, whereas in superconductors it is second order.
To better  clarify  this point, it is fundamental to inquire on the 
temperature dependence of both the penetration depth and coherence length, since
in our approach these two parameters fully determine  the shape of 
the longitudinal chromoelectric field.
\begin{figure}[tb] 
\centering
\includegraphics[scale=0.6,clip]{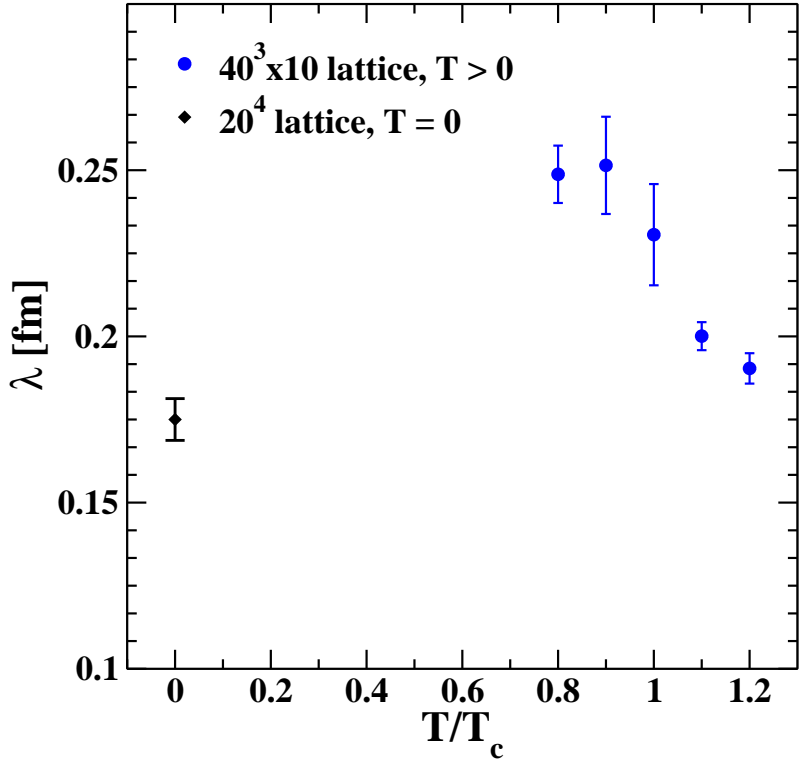}
\caption{(color online). London penetration depth $\lambda$ {\it vs} $T/T_c$. 
The $\lambda_{T=0}=0.1750(63)$ value is included.}
\label{lamdadepth}
\end{figure}
\begin{figure}[tb] 
\centering
\includegraphics[scale=0.6,clip]{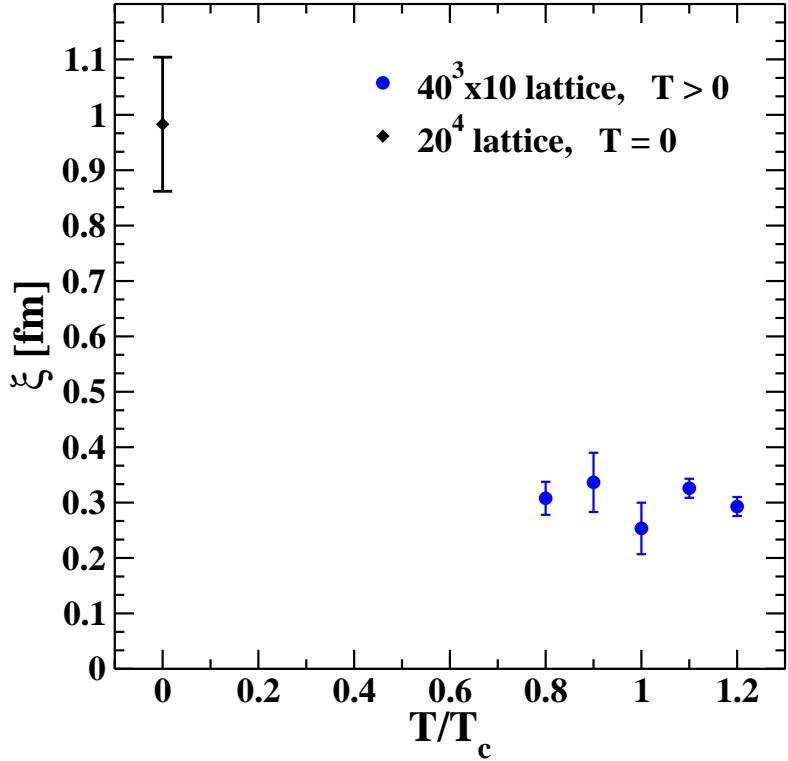}
\caption{(color online). Coherence length $\xi$ {\it vs} $T/T_c$. 
The $\xi_{T=0}=0.983(121)$ value is included.}
\label{coherencelength}
\end{figure}
In Figs.~\ref{lamdadepth} and~\ref{coherencelength} we display the penetration 
depth and the coherence length, in physical units, respectively {\it versus} 
the reduced temperature $T/T_c$. We also report the values of these 
lengths at zero temperature, as previously obtained on  
$20^4$ lattices~\cite{Cea:2013oba,Cea:2014uja,Cea:2014hma}.
As concerns the London penetration length, Fig.~\ref{lamdadepth} shows 
that it seems to slightly increase with respect to the zero-temperature value
for temperatures $T < T_c$, and then to decrease above the critical temperature.
However, the overall variation of $\lambda$ is rather modest, so that, we can 
safely affirm that the London penetration length is almost temperature 
independent. On the other hand, at finite temperatures the coherence length 
suffers from a rather drastic reduction with respect to the zero-temperature 
value. After that, we see from Fig.~\ref{coherencelength} that $\xi$ is almost 
constant across deconfinement. In any case, these 
results indicate clearly that the flux tube survives even after the color 
deconfinement transition. 
\section{Magnetic sector in the deconfined phase}
\label{magnetic}
In this Section we would like to investigate the structure of QCD in the 
high-temperature regime~\cite{Kalashnikov:1984,Nieto:1996pi}. 
At high temperatures, through dimensional reduction, QCD
can be reformulated as an effective three-dimensional theory with the scale of 
the effective couplings given in terms of the temperature. However, the QCD 
effective theory is quite complicated even at high temperatures,
since straightforward perturbation theory does not work due to the presence of  
infrared singularities in the magnetic sector. These nonperturbative effects 
will manifest themselves in correlation functions for the spatial components of 
gauge fields. In fact, it is known since long that gauge-invariant 
correlation functions for the spatial components of gauge 
fields, {\it i.e.} spatial Wilson loops, obey an area law behavior 
in the high-temperature phase, with a nonzero spatial string tension
$\sigma_s$~\cite{PhysRevLett.71.3059,Karsch:1994af}. 
An analysis of the temperature dependence of the spatial string tension thus 
yields information on the importance of the nonstatic sector for 
long-distance properties of high-temperature QCD. It turns out that, for 
temperatures larger than $2 T_c$, the spatial string tension is consistent 
with the behavior
\begin{equation}
\label{spatial}
\sqrt{\sigma_s}  \; = \; \gamma \; g(T) \; T  \; ,
\end{equation}
where $g(T)$ is the temperature-dependent coupling constant, running 
according to the two-loop $\beta$-function, and $\gamma$ is a constant, 
with $\gamma = 0.586 \pm 0.045$ for SU(3)~\cite{Karsch:1994af}, and
$\gamma = 0.369 \pm 0.015$ for SU(2)~\cite{PhysRevLett.71.3059}.  \\ 
We see, thus, that for a better understanding of the nonperturbative structure 
of QCD at high temperature, it is fundamental to arrive at a quantitative 
description of the properties of the spatial string tension.
To this end, we considered the connected correlator built with gauge 
links belonging to the spatial sublattice. Obviously, in this case the 
field strength tensor Eq.~(\ref{fieldstrengthW}) corresponds to the 
chromomagnetic field. As in the previous study, we performed simulations on 
$40^3\times 10$ lattices for physical temperatures ranging from $0.8\, T_c$ 
up to $1.5 \, T_c$. We chose squared Wilson loops with side 
$\Delta \simeq 0.76$ fm (see Table~\ref{tab:fit_summary_2} for the 
summary of our simulation setup). 
\begin{table}[h]
\begin{center} 
\caption{Simulation parameters for the lattice $L_s\times L_t = 40^3\times 10$, 
fitted values of the parameters, and reduced chi-square (chromomagnetic 
sector).}
\label{tab:fit_summary_2}
\begin{tabular}{|c|c|c|c|c|c|c|c|}
\hline\hline
$\beta$ & $\Delta$ [fm] & $T/T_c$ & $\phi$ & $\mu$ & $\xi_v$ & $\chi_r^2$ \\ \hline
6.050	&0.761	&0.8	&7.600(14)	&0.653(5)	&3.313(6)	&1.52\\
6.125	&0.761	&0.9	&8.164(7)	&0.593(3)	&5.978(38)	&2.90\\
6.200	&0.756	&1.0	&7.887(11)	&0.544(4)	&6.413(76)	&1.27\\
6.265	&0.757	&1.1	&8.085(12)	&0.498(6)	&7.572(117)	&0.45\\
6.490	&0.759	&1.5	&9.475(80)	&0.393(23)	&10.793(721)	&0.01\\
\hline\hline 
\end{tabular} 
\end{center}
\end{table}
\begin{figure}[tb] 
\centering
\includegraphics[scale=0.6,clip]{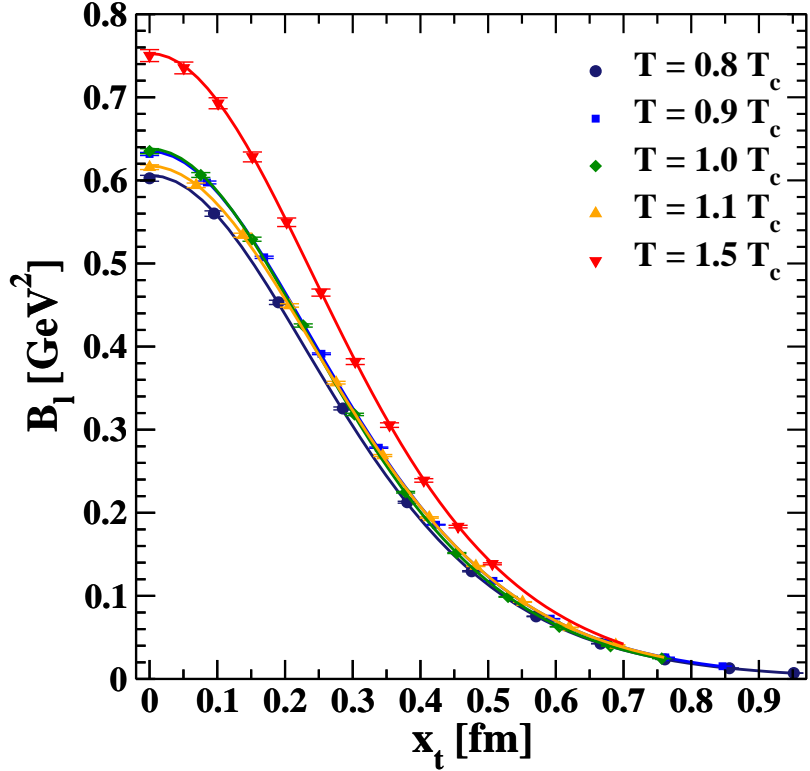}
\caption{(color online). Transverse profile of the longitudinal chromomagnetic 
field {\it vs} the transverse distance across the deconfinement temperature.
The solid lines are the fit of our data to Eq.~(\ref{clem2}).}
\label{fig:B_vs_xt_phys}
\end{figure}
Remarkably, we found that even in this case the chromomagnetic flux tube is 
built from the longitudinal chromomagnetic field only. Moreover, the 
longitudinal chromomagnetic field profile in the transverse directions is 
accounted for by the function given in Eq.~(\ref{clem2}). 
In Table~\ref{tab:fit_summary_2}, we report the values of the fitted parameters 
together with the reduced chi-square. The transverse profiles of the 
longitudinal chromomagnetic field for different temperatures are displayed in 
Fig.~\ref{fig:B_vs_xt_phys}. Unlike the longitudinal chromoelectric field, we 
see that the strength of the longitudinal chromomagnetic field and the size of 
the flux tube increase with the temperature. Moreover, it turns out that the 
temperature behavior of the chromomagnetic flux tube is consistent with
the observed increase of the spatial string tension. In fact, we evaluated the 
spatial string tension as reconstructed from the profile of the 
chromomagnetic field according to 
Refs.~\cite{Cea:2013oba,Cea:2014uja,Cea:2014hma}
and found results which are in agreement with the direct, standard 
determination from spatial Wilson loops, Eq.~(\ref{spatial}) for SU(3), as 
reported in Ref.~\cite{Karsch:1994af}.
\section{Summary and Conclusions}
\label{conclusions}
In this paper we studied the color field distribution between a static 
quark-antiquark pair in the SU(3) pure gauge theory at finite temperatures. 
To our knowledge, this kind of investigation within the dual 
superconductivity approach is so far unique, after
the pioneer study of Ref.~\cite{DiGiacomo:1990hc}, except for the
preliminary analyses of Refs.~\cite{Shibata:2014spa,Shibata:2014tpa,Shibata:2015vua,Shibata:2015ywl,Kondo:2015qfj}.

For the chromoelectric sector we adopted the connected correlator built with 
Polyakov lines, while for the chromomagnetic sector we used the connected 
correlator built with Wilson loops. We have made use of the publicly available 
MILC code~\cite{MILC}, which has been suitably modified by us in order to 
introduce the relevant observables.  
Indeed, the use of the MILC code permits to carry out
simulations on lattices with considerable spatial and temporal extensions. \\
From previous studies, it is known that, at zero temperature, the 
chromoelectric field generated by a static quark-antiquark pair can be 
described within the dual superconductor mechanism for confinement.
In particular, it has been shown that the transverse profile of the 
longitudinal chromoelectric field can be accurately accounted for by the 
phenomenological functional form given in Eq.~(\ref{clem2}). Remarkably, 
in the present study we found that this last result extends also at finite 
temperatures. Moreover, we found that the flux tube structure survives
to the deconfinement transition. However, the behavior of the flux-tube 
chromoelectric field across the deconfinement transition does not match the 
dual version of the effective Ginzburg-Landau description of ordinary 
type-II superconductors.
In particular, the Ginzburg-Landau parameter $\kappa$ is seen to be 
$\kappa \ll 1$ at zero temperature, while $\kappa \simeq 1$ near the 
deconfinement critical temperature. Indeed, we found that as the temperature is 
increased towards and above the deconfinement temperature $T_c$, the amplitude 
of the field inside the flux tube gets smaller, while the shape of the flux 
tube does not vary appreciably across the deconfinement temperature, thus
leading to a scenario which resembles an ``evaporation'' of the flux tube.

Since our results are quite surprising, some comments are in order here:
\begin{itemize}
\item To exclude the possibility of contamination of our Monte Carlo ensembles
above $T_c$ from configurations belonging to the confined phase (and 
{\it vice versa}), we looked at histograms of the measured values for the 
longitudinal chromoelectric field, configuration by configuration, at fixed 
distance $x_t = 0$ ({\it i.e.} on the axis connecting the static sources) and 
at zero smearing steps; the distribution of our measurements showed a Gaussian 
shape in all temperature regimes ($T = 0.8\,T_c$, $T=T_c$, $T=1.2\,T_c$).

\item In our continuum scaling analysis we performed simulations, at fixed 
temperature, on lattices of spatial size 24, 32, 40, 48, 64 (results for $L_s 
= 40$ and 64 are explicitly shown in the manuscript), thus providing with
a check of the stability of our results in the thermodynamic limit.

\item Our investigation relies to a large extent on the smearing 
procedure and it would be advisable to check the stability of our results
under changes of the smoothing procedure, which we plan to do in future 
studies. We stress, however, that the behavior under smearing of the 
parameter $\phi$, shown in Fig.~\ref{fig:phi_vs_smear}, is just as
expected: after a number of smearing steps which scales in accordance
with the diffusive nature of the process, $\phi$ reaches a broad
maximum or stabilizes, thus signalling the complete washing out of 
fluctuations at the level of lattice spacing, before fluctuations at 
physical length scales are affected. Moreover, even admitting that the 
smearing procedure introduces some alteration in the shape of the flux tube,
it would be quite unlikely that it would do that in such a clever way to 
get profiles of the chromoelectric field in so nice an agreement with 
the function given in Eq.~\ref{clem2} (see Fig.~\ref{fig:E_vs_T}) and with the 
continuum scaling (see Fig.~\ref{fig:E_vs_beta}).
As a matter of fact, we have monitored the shape of the chromoelectric
field during the whole smearing procedure, and always found that the field 
profile changes very mildly, in an interval of smearing steps ranging 
from 10 up to numbers of order 100.

\item The attenuation of the flux we observe above $T_c$ could be explained 
by screening effects; we have fixed the physical distance between the two 
static sources at $\Delta = 0.76$ fm and, according to 
Ref.~\cite{Kaczmarek:2005zn}, such distance is compatible with the screening 
length from the lattice at the deconfining temperature.

\item A direct comparison between our results and those of 
Refs.~\cite{Shibata:2015ywl,Kondo:2015qfj} is not possible, since in these works the distance between the sources is not specified. There is, however, a 
reasonable qualitative agreement. Indeed, both works agree on the persistence 
beyond the critical temperature of the longitudinal component of the
chromoelectric field. Namely, the rightmost panels in Figs.~3 and~4 
of Ref.~\cite{Shibata:2015ywl} show that the chromoelectric field, though 
attenuated in amplitude, survives at values of $\beta$ as large as 6.30, 
corresponding, on a $24^3 \times 6$ lattice, to temperatures of about 
$1.93\,T_c$, much larger than those considered in our work.

\end{itemize} 

We also investigated the chromomagnetic sector which is relevant for the 
QCD effective theory at high temperatures.
We focused on the chromomagnetic flux tube which is responsible for the 
nonzero spatial string tension. Even in the chromomagnetic sector we 
found that the flux tube is built mainly from the longitudinal chromomagnetic
field. Our results showed that the strength and the size of the chromomagnetic 
flux tube increase with the temperature, consistently with the temperature 
behavior of the spatial string tension. Our findings confirm the importance of
long-range chromomagnetic correlations in high-temperature QCD. \\
Finally, it is worthwhile to stress that our results could have important 
phenomenological applications in hadron physics. In particular, we believe that 
they are relevant to clarify the nature
of the initial state of the quark-gluon plasma in heavy-ion collisions.
However, before attempting phenomenological applications, it is important to 
extend the present study to full QCD, {\it i.e.} to the SU(3) lattice 
gauge theory with improved gauge action and dynamical quarks with masses at 
(almost) the physical point.
\section*{Acknowledgments}
This work was in part based on the MILC collaboration's public lattice gauge 
theory code. See {\url{http://physics.utah.edu/~detar/milc}}.
This work has been partially supported by the INFN SUMA project.
Simulations have been performed on BlueGene/Q at CINECA 
(Project  INF14\_npqcd), on the BC$^2$S cluster in Bari, and on the CSNIV 
Zefiro cluster in Pisa.

%\bibliography{qcd}
\providecommand{\href}[2]{#2}\begingroup\raggedright\endgroup

\end{document}